\documentclass[12pt]{revtex4}
\begin{document}
\title { Construction of a more complete quantum fluid model from Wigner-Boltzmann Equation with all higher order quantum corrections }
\author{Anirban Bose$^1$ and Mylavarapu.S. Janaki$^2$}
\affiliation { $^1$ Serampore College, Serampore, Hooghly.\\
$^2$Saha Institute of Nuclear Physics, I/AF Bidhannagar,\\
Calcutta 700 064, India}

\begin{abstract}
A semiclassical Quantum Hydrodynamic model has been derived by taking the moments of the Wigner-Boltzmann equation. For the first time, the closure has been achieved by the use of the momentum shifted version of all order quantum corrected solution of the Wigner-Boltzmann equation and that has considerably extended the applicability of the model towards the low temperature and high density limit. In this context, the importance of the correlation and exchange effects have been retained through the Kohn-Sham equation in the construction of the Wigner-Boltzmann equation.  The validity of the approach is subject to the existence of the Taylor's expansion of the associated Kohn-Sham potential. 
\end{abstract}
\pacs{67.10.Jn,05.30.-d}
\maketitle

\newpage
\section{Introduction}
The theory of many body system can be broadly classified in two categories. First one is the microscopic view where each member is treated individually. Their placement in phase space, their mutual interactions, their micro fields are brought under the focus. In the second method, the identity of an individual becomes unimportant and the particles are treated as a collection. The properties of the system are obtained by averaging over the micro properties of the particles and are reflected through some directly measurable quantities like pressure, temperature as a function of space and time. The microscopic nature is investigated by kinetic equations like the many particle Liouville's equation. On the other hand, the macroscopic nature is revealed through the hydrodynamic equations obtained by taking the moments of the kinetic equations like the Vlasov equation. The choice of the method entirely lies on the nature of the problem under investigation.

The formulation of quantum hydrodynamics attracted  considerable  attention in the past\cite{kn:wya}. The first attempt in this direction is concerned with the quantum version of particle and momentum conservation equations which have been constructed from the Schr$\ddot{o}$dinger equation by expressing the wave function in terms of a  density dependent amplitude and a phase factor. In this method\cite{kn:mad,kn:bo1,kn:bo2} the quantum effect is revealed through the Bohm potential which does not have any classical counterpart. Another approach\cite{kn:tak} is related to the phase space analysis which is connected with the Wigner function and the  moments of quantum kinetic equations. This route has been extensively explored to develop quantum hydrodynamic equations\cite{kn:iaf,kn:fre,kn:mar}. This technique gives rise to an infinite hierarchy of moment equations where any member of the group is connected to the next order moment. Therefore, in order to maintain the consistency with the physical condition a closure technique has to be devised. In this context, Gardner\cite{kn:gard} performed the  moment expansion of
the Wigner-Boltzmann equation near thermal equilibrium and high temperature limit to derive the QHD model with a $h^{2}$ order quantum corrected stress tensor \cite{kn:gard,kn:anco,kn:grub}. In the next stage, QHD is obtained using the entropy extremization technique\cite{kn:jung,kn:dego}. Later this principle is applied to the system of identical particles with a newly defined quantum entropy in terms of reduced density matrix where the indistinguishibilty factor is incorporated\cite{kn:trov}. Application of Grad's moment method\cite{kn:cai}, in the construction of hydrodynamic model, has been also reported.     In this article we have followed the approach of Gardner and are able to introduce the quantum correction to all orders to extend the range of applicability of the concerned hydrodynamic model. In addition to that, the quantum exchange  and correlation effects which are absolutely necessary in this context, have been taken care of through the Kohn-Sham potential\cite{kn:kohn} which converts a many body interacting system to a non-interacting one-particle system with an effective potential.  The conventional Wigner equation deals with the Schr$\ddot{o}$dinger equation of wave mechanics. In order to incorporate the effect of correlation and exchange we have replaced the Schr$\ddot{o}$dinger equation with the Kohn -Sham equation which is a nonlinear version of the former.  
To calculate the average values in the first three moment equations, we have used the all order semiclassical equilibrium solution of the Wigner-Moyal equation\cite{kn:bose}. Quantum Hydrodynamics has a wide range of applicability in the context of semiconductor\cite{kn:gard,kn:manf} physics, thin metal films\cite{kn:cro} and even in the astrophysical systems\cite{kn:bro,kn:has} and in nuclear physics\cite{kn:wal}. Similar models have been proposed and employed for analyzing the thermistor\cite{kn:xie} theory and superfluidity\cite{kn:loff}.

\section{Derivation of the hydrodynamic model}
The Kohn-Sham equation is the Schrodinger equation of the fictitious non-interacting particles where the potential is replaced by the effective external potential (Kohn-Sham potential) to incorporate the correlation and exchange effects \cite{kn:kohn}
\begin{eqnarray}
i\hbar\frac{\partial \psi}{\partial t•}=-\frac{\hbar^{2}}{2m•}\nabla^{2} \psi +\phi_{kh} \psi
\label{v1}\end{eqnarray}
where $\phi_{kh}$ is the Kohn-Sham potential.

The  corresponding Wigner equation is
\begin{eqnarray}
&&\frac{\partial f}{\partial t}+\frac{p}{m}
\frac{\partial f}{\partial {x}}-\frac{\partial
\phi_{kh}}{\partial {x}}\frac{\partial f}{\partial
{p}}+ \sum_{j=1}^{\infty}(-1)^{j+1}C_{j}\hbar^{2j}\frac{\partial^{2j+1}
\phi_{kh}}{\partial x^{2j+1}}
\frac{\partial^{2j+1} f}{\partial p^{2j+1}}=0 \label{v1}\end{eqnarray}

where $f(x,p,t)$ is the single particle quasi-distribution function,

 $$C_{j}=1/{(2)}^{2j}(2j+1)!$$

The above equation is written in a normalized form, where we have defined the following normalized variables.
$$t=\frac{t}{l\sqrt{m\beta}}$$
$$x=\frac{x•}{l•}$$ 
$$p=\frac{p\sqrt{\beta}}{\sqrt{m}}$$
where $l$ and $l\sqrt{m\beta}$ are the length and time scale of the system and $\beta$ is the Boltzman constant.
The normalized equation is obtained as
\begin{eqnarray}
&&\frac{\partial f}{\partial t}+p
\frac{\partial f}{\partial {x}}-\frac{\partial
\phi_{kh}}{\partial {x}}\frac{\partial f}{\partial
{p}}+ \sum_{j=1}^{\infty}(-1)^{j+1}C_{j}\Lambda^{2j}\frac{\partial^{2j+1}
\phi_{kh}}{\partial x^{2j+1}}
\frac{\partial^{2j+1} f}{\partial p^{2j+1}}=0 \label{vv1}\end{eqnarray}
where $$\Lambda =\sqrt{\frac{\hbar^{2}\beta}{ml^{2}}}$$ is the small expansion parameter. This equation is  valid only if
it is possible to develop the potential energy $\phi_{kh}$ in a Taylor series\cite{kn:wig}.
To obtain the hydrodynamic equation, we need to take the moments of the Wigner equation.
The first three moment equations are
\begin{eqnarray}
\frac{\partial n}{\partial t}+\frac{1}{m}\frac{\partial \langle p\rangle}{\partial x}=0
\label{q111}\end{eqnarray}
\begin{eqnarray}
\frac{\partial \langle p\rangle}{\partial t}+\frac{\partial}{\partial x}\langle\frac{pp}{m}\rangle=-n\frac{\partial \phi_{kh} }{\partial x}
\label{q222}\end{eqnarray}
\begin{eqnarray}
\frac{\partial}{\partial t}\langle\frac{p^{2}}{2m}\rangle+\frac{\partial}{\partial x}\langle\frac{pp^{2}}{2m^{2}}\rangle=-\frac{\langle p\rangle•}{m•}\frac{\partial \phi_{kh} }{\partial x}
\label{q333}\end{eqnarray}

To calculate the average quantities, we have used  the following momentum shifted solution containing all higher order quantum corrections \cite{kn:bose} 
$$f(x,p,t)=\prod_{j>0}W_{2j}$$

\begin{equation}W_{2}=\exp(a_{01}-a_{11}\frac{p^{2}}{2})\label{v3}\end{equation}
\begin{equation}a_{01}=A+B\end{equation}
where,
$$A=-\int\left( 1+\frac{{\Lambda}^{2}}{6}\frac{d^{2}\phi_{kh}}{dx^{2}}\right)^{-\frac{1}{2}}\frac{d\phi_{kh}}{dx}dx$$
$$B=-\frac{{\Lambda}^{2}}{8}\int\left (1+\frac{{\Lambda}^{2}}{6}\frac{d^{2}\phi_{kh}}{dx^{2}}\right)^{-1}\frac{d^{3}\phi_{kh}}{dx^{3}}dx$$
 
 and\begin{equation}a_{11}=(1+\frac{{\Lambda}^{2}}{6}\frac{d^{2}\phi_{kh}}{dx^{2}})^{-\frac{1}{2}}\label{v5}\end{equation}

For $j>1$,
$$W_{2j}=\exp (U_{2j})$$
where
$$U_{2j}=\sum_{i=0}^{j}\Lambda^{2j} a_{ij}({\frac{p}{\sqrt{2}}})^{2i}$$
$a_{ij}$ can be obtained from the following equations\begin{equation}\frac{1}{2^{i}}\frac{da_{ij}}{dx}=\frac{i+1}{2^{i}}\frac{d\phi_{kh}}{dx}a_{i+1,j}+b_{ij}\label{v15nn}\end{equation}

where

\begin{equation}b_{ij}=-\frac{\partial^{2j+1} \phi_{kh}}{\partial x^{2j+1}}D_{i}C_{j}\label{v16}\end{equation}
in which, $D_{i}=$ coefficient of $p^{2i+1}$ of the Hermite polynomial $He_{2j+1}(p)$

These equations are true for i=0 to i=j-1

For i=j,

\begin{equation}\frac{1}{2^{i}}\frac{\partial a_{ij}}{\partial x}=-\frac{\partial^{2j+1} \phi_{kh}}{\partial x^{2j+1}}C_{j}D_{j}\label{v17}\end{equation}
This is a first order equation and can be easily solved to obtain the value of $a_{jj}$. Using this result, all the equations of this group can be successively solved to find out the remaining coefficients. This process has been illustrated in \cite{kn:bose}.

Expanding the exponential factors of $W_{2}$, we can write

$$a_{01}=\phi_{kh} +\eta$$
and
$$-a_{11}\frac{p^{2}}{2}=-p^{2}/2 +\theta p^{2}$$
$\eta$ and $\theta$  are containing all higher order terms obtained from $a_{01}$ and $a_{11}$.

Now,
$$f(x,p)=Aexp(-\frac{p^{_{'}2}}{2}+\phi_{kh})exp(\eta +\theta p^{_{'}2})\prod_{j>1}^{} exp[\Lambda^{2j}(\sum_{i=0}^{j} a_{ij}\frac{p^{_{'}2i}}{2^{i}})]$$
which can be written in the following form
\begin{equation}f(x,p)=Aexp(-\frac{p^{_{'}2}}{2}+\phi_{kh})exp(\eta)G\label{q}\end{equation}

and
\begin{eqnarray}
G=\sum_{n} \frac{\theta^{n}}{n!•}p^{_{'}2n}\frac{\Lambda^ { 2\sum_{j>1}j \sum_{i=0}^{j}n_{ij}}\prod_{j>1}\prod_{i=0}^{j}a_{ij}^{n_{ij}}}{2^ { \sum_{j>1} \sum_{i=0}^{j}in_{ij}}\prod_{j>1}\prod_{i=0}^{j}n_{ij}!}p^ {_{'} 2 \sum_{j>1} \sum_{i=0}^{j}i n_{ij}}
\end{eqnarray}
where exponential factors have been expanded as

$$exp(\theta p^{_{'}2})=\sum_{n}\frac{(\theta p^{_{'}2})^{n}•}{n!•}$$
$$ exp[\Lambda^{2j}( a_{ij}\frac{p^{_{'}2i}}{2^{i}})]=\sum_{n_{ij}}\frac{(\Lambda^{2j} a_{ij}p^{_{'}2i})^{n_{ij}}}{2^{in_{ij}}n_{ij}!}$$
n and $n_{ij}$ are positive integers.
and $A(x,t)$ is a slow varying function of x and t and
$${p}^{_{'}}={p}-m{u}$$

${u}$ and ${p}$ are the macroscopic fluid velocity and the molecular momentum respectively. 
Therefore, we can write
\begin{eqnarray}
<pp/m>=u<p>-P
\label{q11}\end{eqnarray}
where
\begin{eqnarray}
<p^{_{'}}p^{_{'}}/m>=-P
\label{v1}\end{eqnarray}
 and
 \begin{eqnarray}
<pp^{2}/2m^{2}>=uW-uP+q
\label{q22}\end{eqnarray}
where P and W are stress tensor and energy density respectively.
\begin{eqnarray}
q=<p^{_{'}}p^{_{'}2}/2m^{2}>
\label{v1}\end{eqnarray}
is the heat flux term and carries important effects of higher moments.
\begin{eqnarray}
W=<p^{2}/2m>=mnu^{2}/2 + <p^{_{'}2}/2m>
\label{v1}\end{eqnarray}

Using eq.(\ref{q}) we obtain
\begin{eqnarray}
<p>=mnu
\label{v1}\end{eqnarray}

where
\begin{eqnarray}
n(x,t)=\int f(x,p)dp
\label{v1}\end{eqnarray}
Performing the momentum  integral with the help of the following identity,
$$\int_{0}^{\infty}x^{2n}e^{-\frac{x^{2}}{a^{2}•}}dx=\sqrt{\pi•}\frac{(2n-1)!!}{2^{2n+1}}a^{2n+1}$$
we obtain  
\begin{eqnarray}
n(x,t)=A\sqrt{2\pi}exp(\eta)\sum_{n} \frac{\theta^{n}}{n!•}\frac{\Lambda^ { 2\sum_{j>1}j \sum_{i=0}^{j}n_{ij}}\prod_{j>1}\prod_{i=0}^{j}a_{ij}^{n_{ij}}}{2^ { \sum_{j>1} \sum_{i=0}^{j}in_{ij}}\prod_{j>1}\prod_{i=0}^{j}n_{ij}!}\frac{\sqrt{m}}{\sqrt{\beta}•}(2n+2 \sum_{j>1} \sum_{i=0}^{j}i n_{ij}-1)!!
\end{eqnarray}

Similarly,
\begin{eqnarray}
P=A\sqrt{2\pi}exp(\eta)\sum_{n} \frac{\theta^{n}}{n!•}\frac{\Lambda^ { 2\sum_{j>1}j \sum_{i=0}^{j}n_{ij}}\prod_{j>1}\prod_{i=0}^{j}a_{ij}^{n_{ij}}}{2^ { \sum_{j>1} \sum_{i=0}^{j}in_{ij}}\prod_{j>1}\prod_{i=0}^{j}n_{ij}!}\frac{\sqrt{m}}{\sqrt{\beta}}(2n+2 \sum_{j>1} \sum_{i=0}^{j}i n_{ij}+1)!!
\end{eqnarray}

We define 
 
 \begin{eqnarray}
C=\sum_{n} \frac{\theta^{n}}{n!•}\frac{\Lambda^ { 2\sum_{j>1}j \sum_{i=0}^{j}n_{ij}}\prod_{j>1}\prod_{i=0}^{j}a_{ij}^{n_{ij}}}{2^ { \sum_{j>1} \sum_{i=0}^{j}in_{ij}}\prod_{j>1}\prod_{i=0}^{j}n_{ij}!}\frac{\sqrt{m}}{\sqrt{\beta}}(2n+2 \sum_{j>1} \sum_{i=0}^{j}i n_{ij}-1)!!
\label{q1}\end{eqnarray}
 \begin{eqnarray}
B=\sum_{n} \frac{\theta^{n}}{n!•}\frac{\Lambda^ { 2\sum_{j>1}j \sum_{i=0}^{j}n_{ij}}\prod_{j>1}\prod_{i=0}^{j}a_{ij}^{n_{ij}}}{2^ { \sum_{j>1} \sum_{i=0}^{j}in_{ij}}\prod_{j>1}\prod_{i=0}^{j}n_{ij}!}\frac{\sqrt{m}}{\sqrt{\beta}}(2n+2 \sum_{j>1} \sum_{i=0}^{j}i n_{ij}+1)!!
\label{q2}\end{eqnarray}
Therefore
\begin{eqnarray}
<p>=mnu
\label{q3}\end{eqnarray}
\begin{eqnarray}
P=nB/C
\label{q4}\end{eqnarray}
\begin{eqnarray}
W=mnu^{2}/2-nB/2C
\label{q5}\end{eqnarray}
Inserting $<p>$, P and W in (\ref{q11}) and (\ref{q22}) 
\begin{eqnarray}
<pp/2m>=(mnu^{2}-nB/C)
\label{q6}\end{eqnarray}
\begin{eqnarray}
<pp^{2}/2m^{2}>=u(mnu^{2}/2-3nB/2C) +q
\label{q7}\end{eqnarray}
As pointed out by Gardner we have allowed a heat flux term $q=-\kappa\nabla T$, even though our distribution function does not allow it. 
Eq.(\ref{q111})-(\ref{q333}) constitute the set of hydrodynamic equations where $<p>$,$<pp/2m>$ and $<pp^{2}/2m^{2}>$ are obtained from eq.(\ref{q3}), (\ref{q6}) and (\ref{q7}) respectively.

\section{Conclusion}

In this article we have established the Quantum Hydrodynamic model that incorporates  quantum corrections to all orders. To do this we have followed the procedure introduced by Gardner who has developed the O($h^{2}$) fluid model by taking the moments of the Wigner-Boltzman equation to obtain the particle, momentum and energy conservation equations. In this process, Gardner has used the local "momentum-shifted" version of the O($h^{2}$) solution of Wigner equation with slow space and time dependence of density, temperature and macroscopic velocity to calculate the associated average values with the expectation that the equations will be valid, as in classical hydrodynamics, to describe systems even not near thermal equilibrium condition. Here, we have introduced  the "momentum-shifted" version solution containing all higher order quantum corrections. Consequently, all higher order quantum terms are introduced to the stress tensor and the energy density and the range of validity of the QHD model has been extended considerably towards the high density and low temperature limit. In this connection, it is important to include the exchange and correlation effects of the particles and that has been taken care of through the application of the Kohn-Sham equation which has replaced the Schrodinger equation in the construction of the Wigner equation. This should be emphasized here that as we consider the higher order quantum terms, the correction due to the exchange and correlation should become more and more important and that implies the necessity of considering those effects when we are interested in the Quantum Hydrodynamic model with all order correction terms. The condition of validity of  this model is restricted to the existence of the proper Taylor expansion of the potential function of the Kohn-Sham equation.  

At the end, it should be mentioned that in this model the distribution used to calculate the average quantities is unable to produce the heat flux. One way to get around this problem is the application of the Chapman-Enskog expansion\cite{kn:vin} of the kinetic equation with the BGK type relaxation term at the right hand side. This problem will be taken up as our future work.
  
Finally, in the context of Quantum hydrodynamic model, the importance of this work is the construction of a more complete quantum fluid model with an extended probing capacity for analyzing low temperature and high density systems of different branches of physics.

\end{document}